\documentclass[12pt,preprint]{aastex}
\usepackage{graphicx}

\def\tobjnum{10181\ }
\def\objsnnum{369\ }

\def\sindex{\ifmmode {{\cal S}}\else
                ${S\  }$\fi}
\def\sindexns{\ifmmode {{\cal S}}\else
                ${S}$\fi}

\def\objnum{25\ }
\def\psnum{46\ }
\def\psnumspc{28\ }
\def\starnum{26\ }
\def\objpmnum{360\ }
\def\mdwawfsnum{18\ }

\def\wdnum{8\ }
\def\goodqsonum{2\ }

\def\mb{\ifmmode {{\rm B_{435}}}\else
                ${\rm B_{435}}$\fi}
\def\mv{\ifmmode {{\rm V_{606}}}\else
                ${\rm V_{606}}$\fi}
\def\mi{\ifmmode {{\rm i_{775}}}\else
                ${\rm i_{775}}$\fi}
\def\mz{\ifmmode {{\rm z_{850}}}\else
                ${\rm z_{850}}$\fi}

\begin{document}
\title{Stars in the Hubble Ultra Deep Field}
\author{N. Pirzkal\altaffilmark{1}, K. C. Sahu\altaffilmark{1}, A. Burgasser\altaffilmark{2},   L. A. Moustakas\altaffilmark{1}, C. Xu\altaffilmark{1}, S. Malhotra\altaffilmark{1}, J. E. Rhoads\altaffilmark{1}, A. M. Koekemoer\altaffilmark{1}, E. P. Nelan\altaffilmark{1}, R. A. Windhorst\altaffilmark{3}, N. Panagia\altaffilmark{1}, C. Gronwall\altaffilmark{4}, A. Pasquali\altaffilmark{5},  J. R. Walsh\altaffilmark{6}}

\altaffiltext{1}{Space Telescope Science Institute, 3700 San Martin Drive, Baltimore, MD21218, USA}
\altaffiltext{2}{Department of Astrophysics, Division of Physical
Sciences, American Museum of Natural History, Central Park West at
79$^{th}$ Street, New York, NY 10024, USA; adam@amnh.org}
\altaffiltext{3}{Dept. of Physics \& Astronomy, Arizona State University, Street: Tyler Mall PSF-470, P.O. Box 871504, Tempe, AZ 85287-1504, USA}
\altaffiltext{4}{Department of Astronomy \& Astrophysics, Pennsylvania State University, 525 Davey Laboratory, University Park, PA 16802}
\altaffiltext{5}{Institute of Astronomy, ETH H\"{o}nggerberg, 8093 Zurich, Switzerland}
\altaffiltext{6}{ESO/ST-ECF, Karl-Schwarschild-Strasse 2, D-85748, Garching bei M\"{u}nchen, Germany}

\begin{abstract}
We identified \psnum unresolved source candidates in the  Hubble Ultra Deep Field, down to $\mi=29.5$. Unresolved objects were identified using a parameter \sindexns, which measures the deviation from the curve-of-growth of a point source. Extensive testing of this parameter was carried out, including the effects of decreasing signal-to-noise and of the apparent motions of stars, which demonstrated that stars brighter than $\mi=27.0$ could be robustly identified. Low resolution grism spectra of the \psnumspc objects brighter than $\mi=27.0$ identify \mdwawfsnum M and later stellar type dwarfs, 2 candidate L-dwarfs, \goodqsonum QSOs, and 4 white dwarfs. Using the observed population of dwarfs with spectral type M4 or later, we derive a Galactic disk scale height of $400 \pm 100 {\rm \ pc}$ for  M and L stars. The local white dwarf density is computed to be as high as $( 1.1 \pm 0.3) \times 10^{-2}  {\rm \ stars/pc^3}$. Based on observations taken 73 days apart, we determined that  no object in the field has a proper motion larger than 0.027''/year ($3\sigma$ detection limit). No high velocity white dwarfs were identified in the HUDF, and all four candidates appear more likely to be part of the Galactic thick disk. The lack of detected halo white dwarfs implies that, if the dark matter halo is 12 Gyr old, white dwarfs account for  less than 10\% of the dark matter halo mass. 
\end{abstract}


\section{Introduction}
Studying the stellar content of our galaxy down to the faintest possible magnitudes is necessary to study the structures of the Galactic disk and population II halo (referred to as Galactic halo thereafter) and stellar population \citep{mendez96}. The nature of the Galactic dark matter halo has in addition been the subject of some debate since the MACHO project \citep{alcock2000} observed lensing events in the direction of the Large Magellanic Cloud (LMC) and claimed that 20\% of the dark matter halo mass could be in the form of low mass objects with masses smaller than $1.0 M_\odot$ \citep{alcock2000}. While there are alternative explanations for the observed lensing events \citep[][and references therein]{sahu1994,sahu2003,gates2001}, the detection of faint, blue unresolved objects with high proper motion has been used as evidence for the existence of a large population of old white dwarfs in the Galactic dark matter halo. \citet{oppenheimer2001} estimated that about 2\% of the dark matter halo mass could be accounted for by old halo white dwarfs. There has since been some debate as to whether the former result might have been caused by a dynamically warmer thick disk component \citep{reid2001} and the latter might have been affected by the misidentification of  faint blue extra-galactic sources (\citet{kilic2004}, also see \citet{reid2004} for a complete review).
The existence of ultra cool white dwarfs ($T \leqslant 4000\degr {\rm \ K}$) has however been confirmed in recent SLOAN data \citep{gates2004}.\\
A search for distant white dwarfs using deeper observations provides another opportunity to firmly examine the existence of a significant population of old white dwarfs in the Galactic halo and the dark matter halo. As shown by \cite{kawaler1996},  deep broad band imaging can probe a significant portion of the Galactic dark matter halo and should identify Galactic dark halo white dwarfs in the process.
 Ideally, one would like to identify white dwarf candidates spectroscopically, instead of relying on broadband colors only, and to determine their Galactic disk or halo memberships using sensitive proper motion measurements.\\
The Advanced Camera for Survey (ACS) HST observations of the Hubble Ultra Deep Field \citep[HUDF,][]{beckwith2004} and their unprecedented depths offer a new opportunity to detect faint stars, such as white dwarfs, at high galactic latitude.  Additionally, the GRAPES survey  \citep[GRism ACS Program for Extragalactic Science. PI: Malhotra, see description in ][]{pirzkal2004} provides low resolution spectra of most sources in the HUDF. These spectra can be used to confidently determine spectral types of stars in the HUDF, down to $\mi = 27.0$ mag. More importantly, these spectra allow one to differentiate unresolved, blue, faint extra-galactic objects (e.g. QSO) from white dwarfs and other stars.\\
Section \ref{sindexsec} outlines how unresolved sources were identified in the HUDF, while the spectroscopic identification of unresolved sources is discussed in Section \ref{fitting}.  Section \ref{discussion} describes an analysis of the cumulative number  distribution of  the late-type dwarfs identified in the HUDF and its implication for the Galactic disk structure (Section \ref{mdwarfs}). Section \ref{discussion}  also includes a description of the white dwarf content of the Galactic disk and halo (Section \ref{WD}). 

\section{Description of the data}\label{thedata}
The HUDF was imaged using the ACS and the F435W (\mb), F606W (\mv), F775W (\mi) and F850LP (\mz) bands at different epochs. The observations were carried out by cycling through each filter while using a small dither pattern to mask out the ACS inter-chip gap and other cosmetic problems. 
The HUDF was observed during 4 different visits, rotating the field of view to a different position angle each time \citep[0\degr, 8\degr, 85\degr, 91\degr\ with respect to the first visit,][]{beckwith2004}.
The individual ACS images had an integration time of 1200 seconds, a size of 4096x4096 pixels and a pixel scale of 0.050''/pixel. These images were combined together by drizzling them onto a finer pixel grid with a pixel scale of 0.030''/pixel, while geometrically correcting these images. The limiting AB magnitudes (S/N=10) of the \mi\ and \mz\ images are 29.0 and 28.7 mag, respectively.
In this paper, images taken from 09/24/2003 to 10/28/2003 and from 12/04/2003 to 01/15/2004 are referred to as Epoch 1 and 2 images, respectively. The elapsed time between the mean observing dates of the two epochs is 73 days.  Images from Epoch 1 and Epoch 2 were taken at position angles that are nearly 90\degr\ apart.\\
While the deep, combined ACS HUDF images described above were used to determine which objects are unresolved (Section \ref{sindexsec}), eight special partial image stacks were also assembled to take advantage of the fact that the data were taken over  a period of several months. Four partial image stacks were created for each HUDF Epoch, using the exact same method used to create the original combined HUDF images. These partial image stacks can be used to determine the variability  \citep{cohen2004} as well as measure the proper and parallax motions of the objects in the  field. The eight partial image stacks have the same physical properties as the combined ACS HUDF images but with one eighth of the integration time. Note that the cosmic ray rejection step was done separately while assembling each partial image stack so that any strongly varying or moving object would not be removed by the cosmic ray rejection process.
The astrometric registration of the combined HUDF images and the eight partial image stacks is estimated to be better than 0.1 ACS pixel (0.005'' or 0.16 HUDF image pixel) \citep{beckwith2004}.\\
Spectroscopic observations of the UDF using the ACS slitless grism mode (G800L) were carried out  as part of the GRAPES project. The reader is referred to \citet{pirzkal2004} for an in-depth description of these spectra. These spectra are low resolution spectra (R=100), span a wavelength range of $5500 \AA \leqslant \lambda \leqslant 10500 \AA $,  and can be used to spectroscopically identify objects in the field.\\
The HUDF Public Catalog 1.0, containing \tobjnum objects, was used to identify objects in the field and all objects are referred to by their ID in this catalog (UID). The magnitudes of these objects were re-measured using matched aperture photometry with apertures defined using Sextractor \citep{Sextractor} and the \mi\ band HUDF image. The photometric ACS magnitude zero-points (25.65288, 26.49341, 25.64053, 24.84315 in the \mb, \mv, \mi, \mz{\ } bands respectively) were taken from  the GOODS survey \citep{giavalisco2004} .

\section{Identification of point sources in the HUDF} \label{sindexsec}
We identified unresolved objects in the HUDF using a single-index (\sindexns) method that is based on a simple analysis of the curve of growth of the light distribution of an object. The IRAF task RADPROF was used to produce these light curves of growth. This task first computes the intensity weighted center of the object (in a 3x3 pixel box), and then performs aperture photometry using increasing aperture sizes (up to a radius of 10 pixels). These photometric measurements are then fit (using eight 3 $\sigma$ rejection cycles) to a cubic spline of order 5.  RADPROF was applied to stamp images of objects which were masked using Sextractor segmentation maps. These segmentation maps were used to avoid contamination from neighboring objects. The RADPROF parameters described here were selected because, following extensive initial testing, they were found to result in curves of growth that are not too sensitive to the presence of faint nearby objects while consistently remaining sensitive to the presence of faint, diffuse extended emission around objects. The stellarity \sindex of an object is defined as $$  \sindex^2 = {1 \over{r_{rmax}}} \sum_{r=0}^{r_{max}}  (F_{obj}(r) - F_{psf}(r) )^2 $$ where $F_{obj}(r)$ and  $F_{psf}(r)$ denote the curve of growth of light, normalized to 1.0 at a distance of $r_{max}$, of the object and of a point source. A value of 10. was used for $r_{max}$.
                       
\begin{figure}
\includegraphics[width=5.0in]{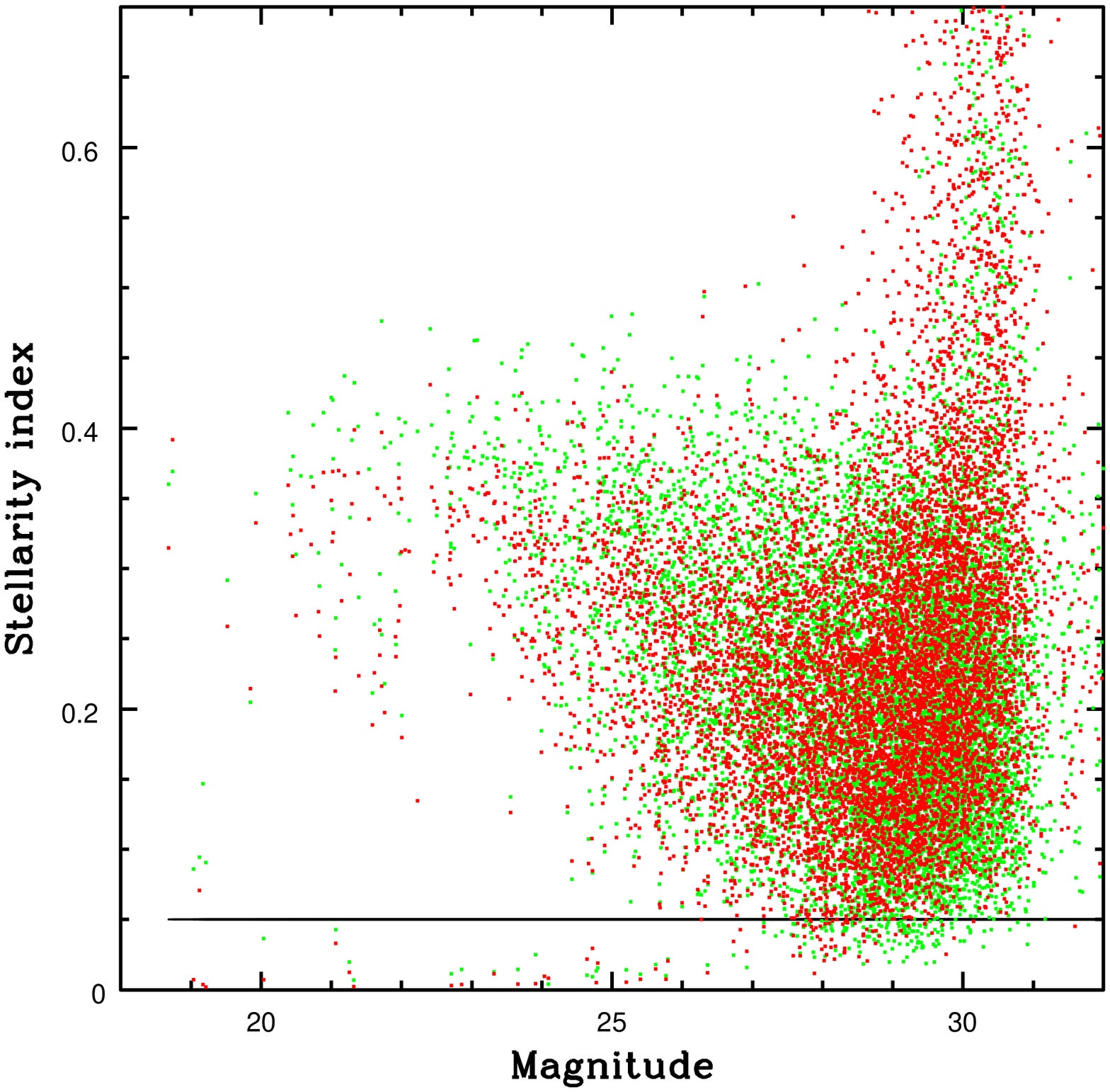}
\caption{The distribution of ${\sindex}_{\mi}$ (green) and ${\sindex}_{\mz}$ (red) values for the \tobjnum HUDF objects. The stellarity index \sindex was defined (see Section \ref{sindexsec}) so that higher values correspond to objects which are increasingly resolved in the images. We selected unresolved objects fainter than $\mi \geqslant 20.0$ mag to be unresolved if  ${\sindex}_{\mz} \leqslant $0.05, and ${\sindex}_{\mz} \leqslant $0.15 for brighter objects which are saturated.\label{fig1}} 
\end{figure}

                       The index \sindex is then simply the square root of the sum of the squared differences between the two curves. Unresolved objects have \sindex values that are close to 0.0 while increasingly resolved objects produce increasingly larger (unbound) values of \sindexns. The detailed calibration of \sindex was done in several stages, each time refining the reference curve of growth that was used. This is described in some detail below.\\
First, a few  obvious, bright, unresolved and unsaturated sources were pre-selected for further study. This was done in both the \mi\  and \mz\  band images separately.  The GRAPES spectra of these sources were examined and  a main-sequence star was selected (UID 911). The curves of growth of this object were measured separately in the \mi\ and \mz\ bands and were used to compute $\sindex_{\mi}$ and $\sindex_{\mz}$ values  for all  \tobjnum objects in the HUDF.  
A new set of \objnum bright point sources, well distributed over the whole ACS field, were then carefully selected based on their low $\sindex_{\mi}$  values (arbitrarily chosen to be ${\sindex_{\mz}} \leqslant 0.04$). The GRAPES spectra of these \objnum objects were visually inspected to ensure that these objects were stars. Stamp images of these objects  were individually drizzled \citep{drizzle} and carefully re-centered to produce a stack of images which were then combined together to produce empirical \mi\ and \mz\  PSF images. A sigma-based rejection algorithm was used to avoid the effect of any faint neighbor in the final combined images.
Final \mi\ and \mz\ band reference curves of growth were generated using these empirical PSF estimates and  all of the $\sindex_{\mi}$ and $\sindex_{\mz}$ values were re-computed. Figure \ref{fig1} shows the distributions of the \sindex values in both the \mi\ and \mz\ bands, as a function of magnitude. There are clearly two distinct groups of object in the HUDF images: one group consists of object that are completely or partially resolved, forming a large distribution around the \sindex value of 0.2 and $\mi=28.0$ mag. A second group consists of unresolved objects with low \sindex values at all magnitudes. Bright stars in Figure \ref{fig1} are observed to have \sindex values which are consistently lower than 0.05. The exception to this are bright stars near $\mi=18$ mag. which are saturated and have artificially high values of \sindex ($\approx 0.1$) .  The effect of bright star saturation on the measured values of \sindex were investigated and values of $0.1 \leqslant \sindex \leqslant 0.15$ are expected if the center of a star is saturated or missing in the HUDF images. Nevertheless, the populations of resolved and unresolved objects appear to be well separated at brighter magnitudes in Figure \ref{fig1}. \\
Unresolved sources in the HUDF were selected using the criterion that ${\sindex}_{\mi} \leqslant  0.05$ mag. Four bright saturated stars were added to the final object list manually. The final list of unresolved UDF sources contains \psnum objects, spanning the magnitude ranges of $19.0 \leqslant \mi \leqslant 29.5$ mag and $18.3 \leqslant \mz  \leqslant 29.2$ mag.  Table \ref{table1} summarizes some of the properties of these objects. 

\subsection{Testing the selection process}\label{test}
\begin{figure}
\includegraphics[width=5.0in]{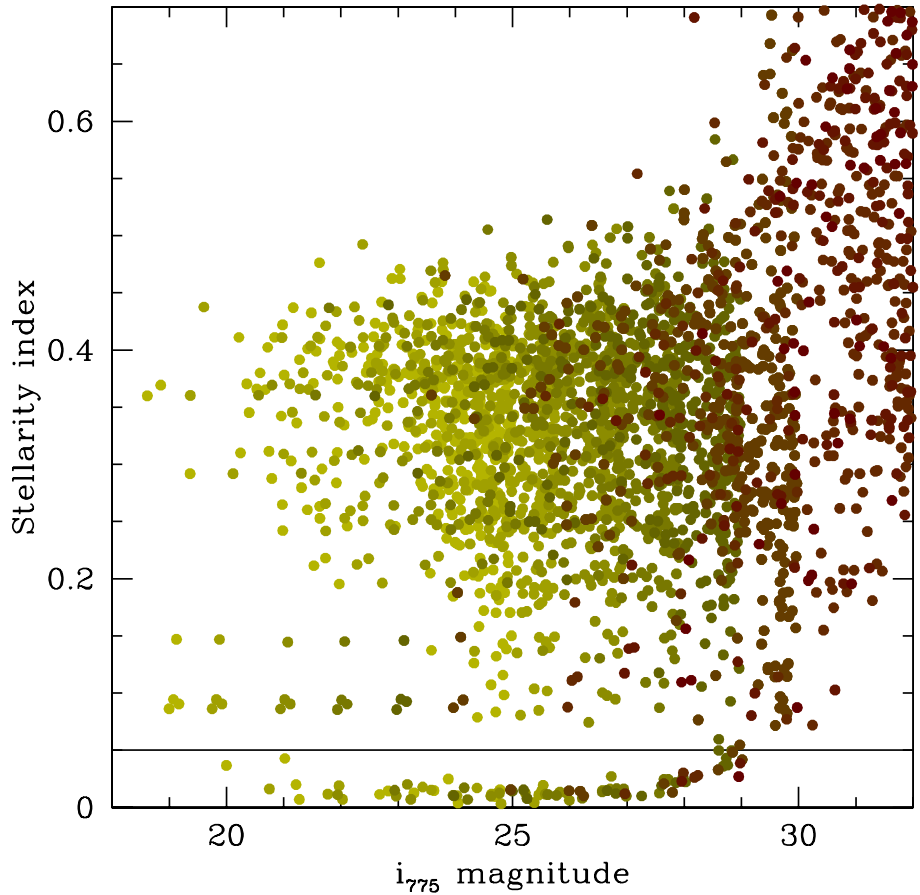}
\caption{The distribution of ${\sindex}_{\mz}$ values for  the brightest \objsnnum objects in the field.  Each object appears 10 times in this figure, with increasingly lower signal to noise (increasingly darker dots).  The observed increase in \sindex values as a function of magnitude at $\mi \geqslant 29.0$ mag is similar to the one observed in Figure \ref{fig1} where the real distribution of the \sindex values from the \tobjnum HUDF objects is shown. This demonstrates the robustness of  \sindex to distinguish between point sources and extended objects down to faint magnitudes.  At fainter magnitudes, images of stars become increasingly dominated by noise which tends to increases the measured values of \sindex. The same test was performed in the \mi\  band and yielded an identical behavior.  \label{fig2}}
\end{figure}

The ability to distinguish between resolved and unresolved objects using  \sindex values was extensively tested,  using a subset of the HUDF object catalog containing the 369 brightest ($\mz \leqslant 25.0$ mag) objects in the field. Simulated \mi\ and \mz\ band images of these objects --- made 0.0, 0.75, 2.0, 3.0, 4.0, 5.0, 6.0, 7.0, 8.0, and 9.0 magnitudes fainter --- were created and \sindex values measured using each simulated image, exactly as described in Section \ref{sindexsec}. The \sindex value of each of these 369 test objects was thus measured 10 times, each measurement corresponding to a different apparent magnitude. Figure \ref{fig2} shows how the \sindex value remains largely unchanged as an object is made fainter and noisier, all the way down to  $\mi=29.0$ mag. The vertical pattern seen in this figure is a result of the 10 magnitude steps used in this simulation.  This test demonstrates a very small fraction of misclassification:  not a single extended object was misclassified as a star (e.g. no object with ${\sindex} \geqslant 0.05$ originally in the HUDF images was measured to have ${\sindex} \leqslant 0.05$ as the object was made fainter), and fewer than 5\% of stars were misclassified down to $\mi=29.0$ mag. While some objects shown in Table \ref{table1} are fainter than this, only objects significantly brighter than this limit ($\mi \leqslant 27.0$ mag) are actually included in the analysis described below and these are expected to be  bona fide unresolved objects.

\subsection{Testing the effect of proper motion}\label{PM}
Some of the stars in the HUDF objects could have a high tangential-velocity ($V_t$) and a correspondingly large proper motion. Such objects would be  blurred in the combined HUDF images and would appear to be resolved.  Similarly, nearby objects would suffer from a significant amount of parallax, again causing these objects to be blurred and mis-identified.  The index \sindex  proved relatively insensitive to a small amount of object motion. This is not unexpected since \sindex was tuned to the detection of  faint extended components, The curves of growth produced by RADPROF were computed by summing, using several sigma rejection steps, the pixel intensities inside annuli of increasing radii. The effect of a nearby source is strongly diminished as its separation from the main source increases and the number of unaffected pixels increases. In addition, the use of a fifth order spline imposes some smoothness on the curve of growth which further lowers the effect of a nearby object.  Still, the effect of object motion on measured values of \sindex was carefully investigated by generating images of the  \psnum unresolved HUDF objects that were selected as described in Section \ref{sindexsec}. These were appropriately shift-and-added, accounting for the orientation and pointing used for each individual HUDF observation while allowing objects to move as a function of time, to simulate the effect of proper motion.  As  larger values of proper motion were introduced, the \sindex values  increased only slightly and  \sindex was found to be mostly unaffected by shifts smaller than 2.5 HUDF pixel ($\mu = 0.3{\rm \ ''/year}$). 70\% of objects were still properly identified as point sources  when shifts 2.5 and 5.0 HUDF pixels were applied ($ 0.3 \leqslant \mu \leqslant 0.6$). 

\section{Identifying stars}\label{fitting}

\subsection{Using Grism spectra to identify stars}\label{stars}
\begin{figure}
\includegraphics[width=7.0in]{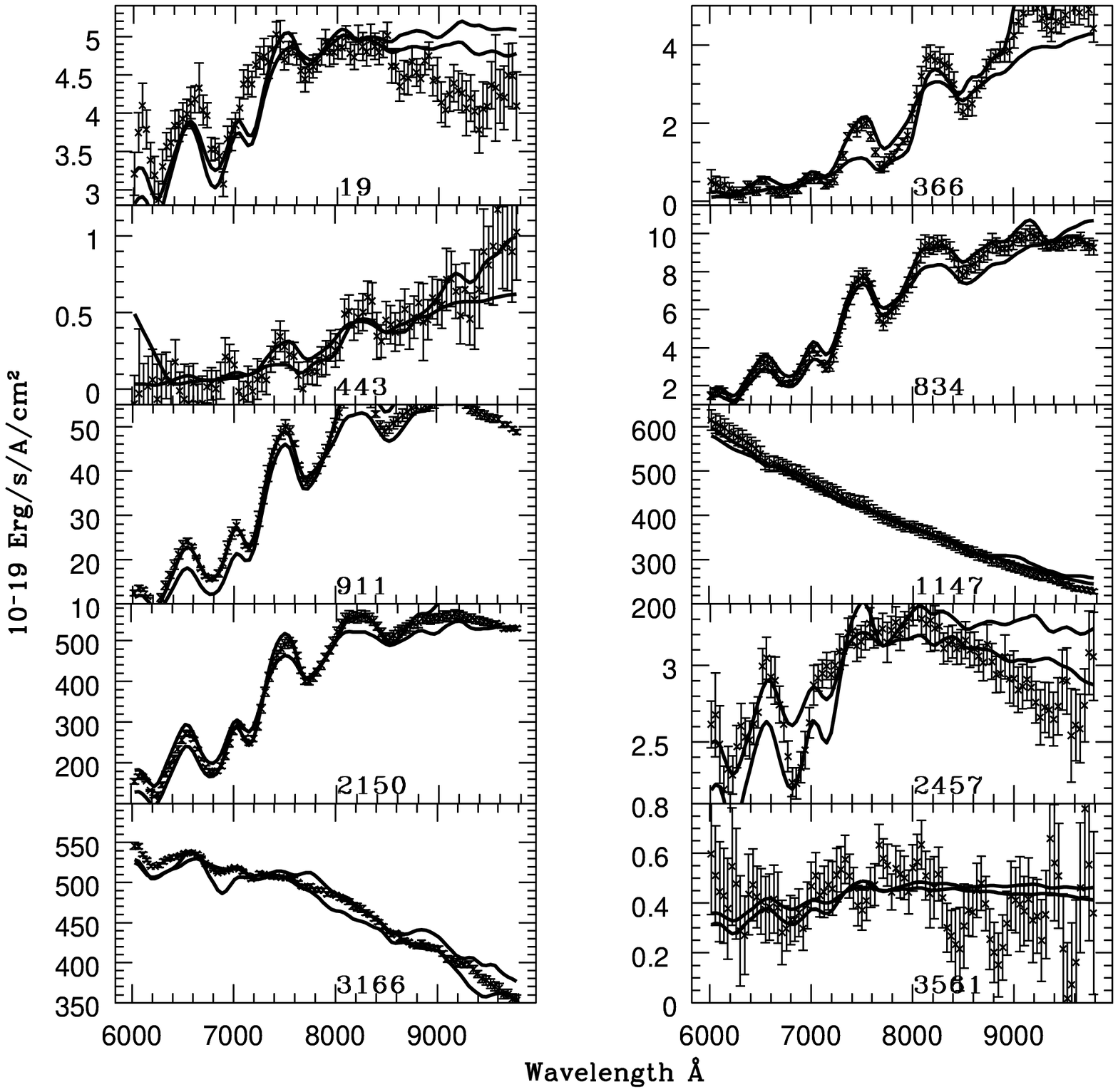}
\caption{GRAPES spectra of unresolved objects in the HUDF and the two best matching templates (solid lines) listed in Table \ref{table2}\label{all1}} 
\end{figure}

\begin{figure}
\includegraphics[width=7.0in]{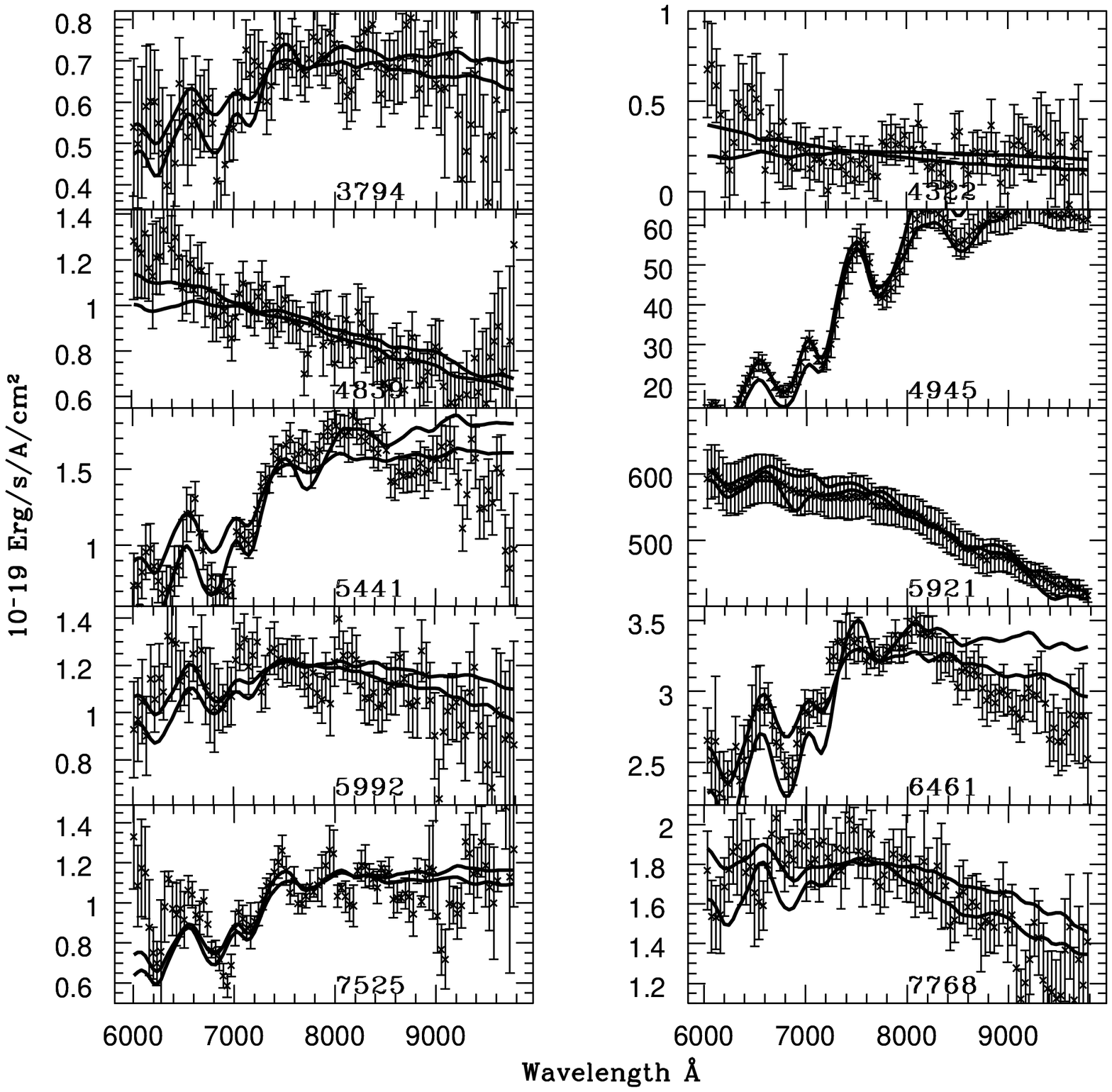}
\caption{GRAPES spectra of unresolved objects in the HUDF and the two best matching templates (solid lines) listed in Table \ref{table2}\label{all2}} 
\end{figure}

\begin{figure}
\includegraphics[width=7.0in]{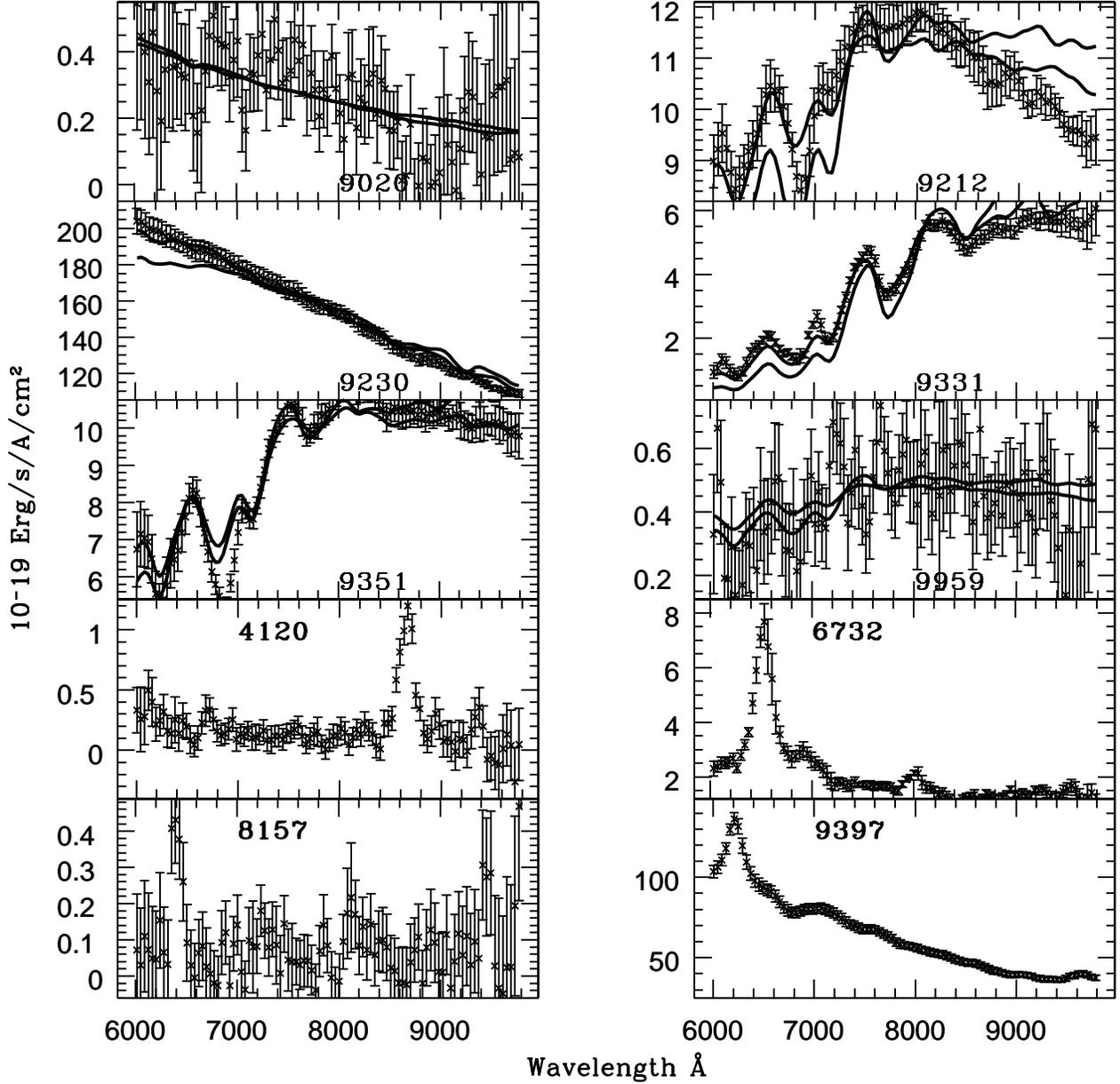}
\caption{GRAPES spectra of unresolved objects in the HUDF and the two best matching templates (solid lines) listed in Table \ref{table2}. The bottom 4 objects are the extra-galactic sources 4120, 6732, 8157, 9397 estimated to be at redshifts of $z=2.1$, $z=3.2$, $z \geqslant 3.0$, $z=3.0$, respectively \citep[][]{chun2004}.\label{all3}} 
\end{figure}

\begin{figure}
\includegraphics[width=5.0in]{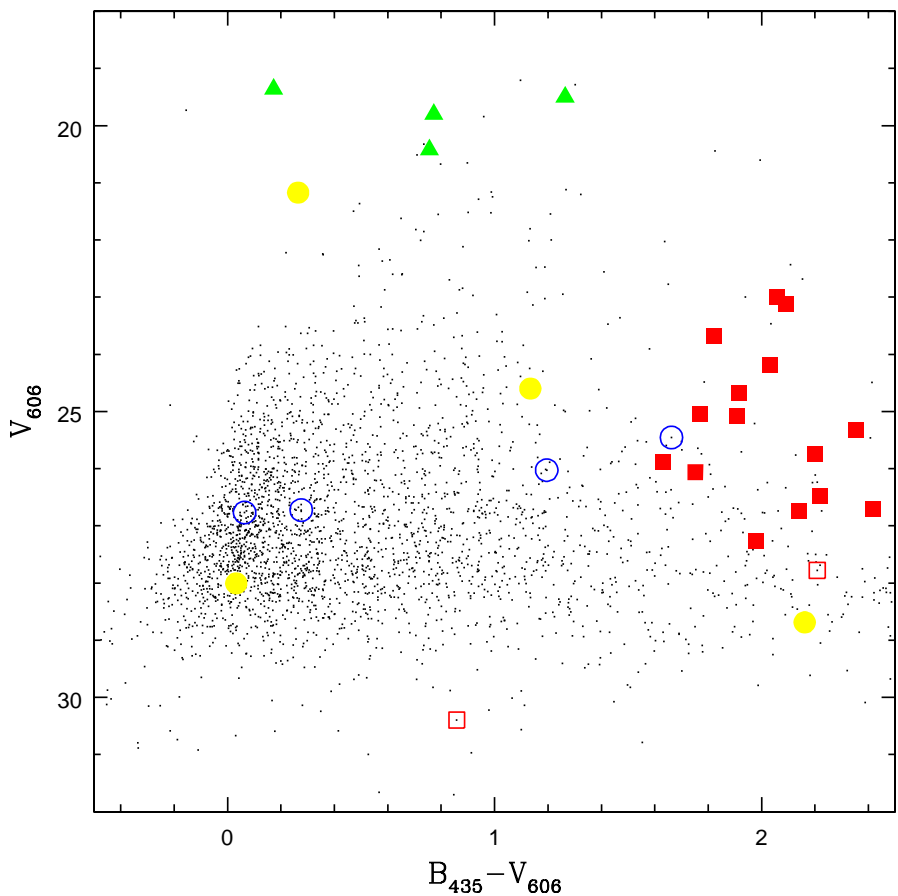}
\caption{The location of the \psnumspc spectroscopically identified sources in a magnitude color plot. The small, dark points are the entire set of objects in the HUDF catalog with $\mi  \leqslant 29.0$ mag and $\mz \leqslant 28.2$ mag.  Solid red squares are M dwarfs. Empty red squares  are the two L dwarf candidates. Solid green triangles represent the non M or L dwarfs stars which could be main-sequence stars. The empty blue circles are white dwarf candidates (Table \ref{table4}).    The solid yellow circles represent the QSOs. \label{VBV}}
\end{figure}

\begin{figure}
\includegraphics[width=5.0in]{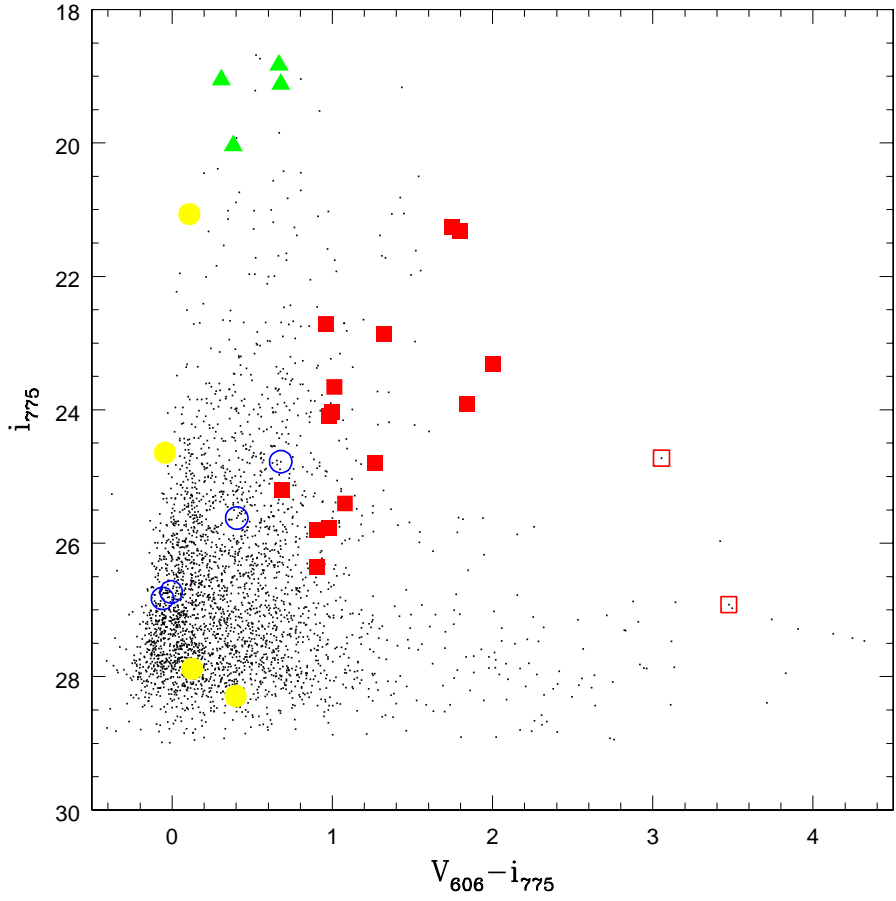}
\caption{The location of the \psnumspc spectroscopically identified sources in a magnitude color plot. The symbols are the same as in Figure \ref{VBV}. \label{IVI}}
\end{figure}

The spectral type  of  \psnumspc  unresolved objects brighter than $\mi=27.0$ could be determined robustly using template fitting of their GRAPES spectra \citep{pirzkal2004}.  The template set included \citet{pickles1998} main-sequence templates,  L and T dwarf spectra \citep{kirkpatrick2000,burgasser2003}, DA and DB white dwarf spectra \citep{harris2003} obtained from the SDSS archive, and some low temperature (3000\degr K, 5000\degr K, and 8000\degr K) model spectra of white dwarfs  \citep{harris2001}. The quality of each spectral fit was evaluated visually and the results of these fits are given in Table \ref{table2}. Objects too faint to have reliable fits are indicated. The GRAPES spectra and the two best-fits listed in Table \ref{table2} are shown in Figures \ref{all1}, \ref{all2}, and \ref{all3}. Unresolved extra-galactic sources are labelled as GAL in Table \ref{table2} and are discussed in more details in Section \ref{QSO}.\\
Of the \psnum point sources detected in the HUDF,  \starnum stars, all  brighter  than  $\mi=27.0$ mag were identified. As shown in Table \ref{table2}, the majority of stars in the HUDF ({\mdwawfsnum} out of  \starnum)  are found to be M dwarfs. Figures \ref{VBV} and \ref{IVI} show these objects in \mb\ vs. \mb-\mv\ and \mi vs. \mi-\mv\ plots. Such plots have in the past shown a lack of faint red objects in the region ${\mv} \geqslant  27.0$ mag  and $(\mb-\mv) \geqslant 1.0$ \citep{mendez96}. Objects found in this region of the plot, appear to be either extra-galactic or L-type dwarf spectra (UID 443, 366). Further NIR observations will be required to verify the nature of these 2 L-dwarf candidates by  confirming that they both have the expected blue NIR colors of this type of object.\\
The HUDF field is at high galactic latitude (b=223\degr, l=--54\degr) and the effects of extinction and reddening are small  \citep[${\rm A_V} =  0.026$,][]{schlegel1998}. The distances to the HUDF stars, which are shown in Table \ref{table2}, were determined using their observed \mv\ magnitudes, their estimated stellar types, and the dwarf star main-sequence absolute Johnson V magnitudes from \citet{allen}. The latter were converted to the \mv\ filter bandpass and AB magnitudes to be consistent with the data at hand.  Distance estimates obtained under the assumption that these objects are white dwarfs with intrinsic magnitudes of $M_{mv}=14.0 {\rm\ and\ } 17.0$ are also given.

\subsection{Proper motion measurements}\label{PMm}
\begin{figure}
\includegraphics[width=7.0in]{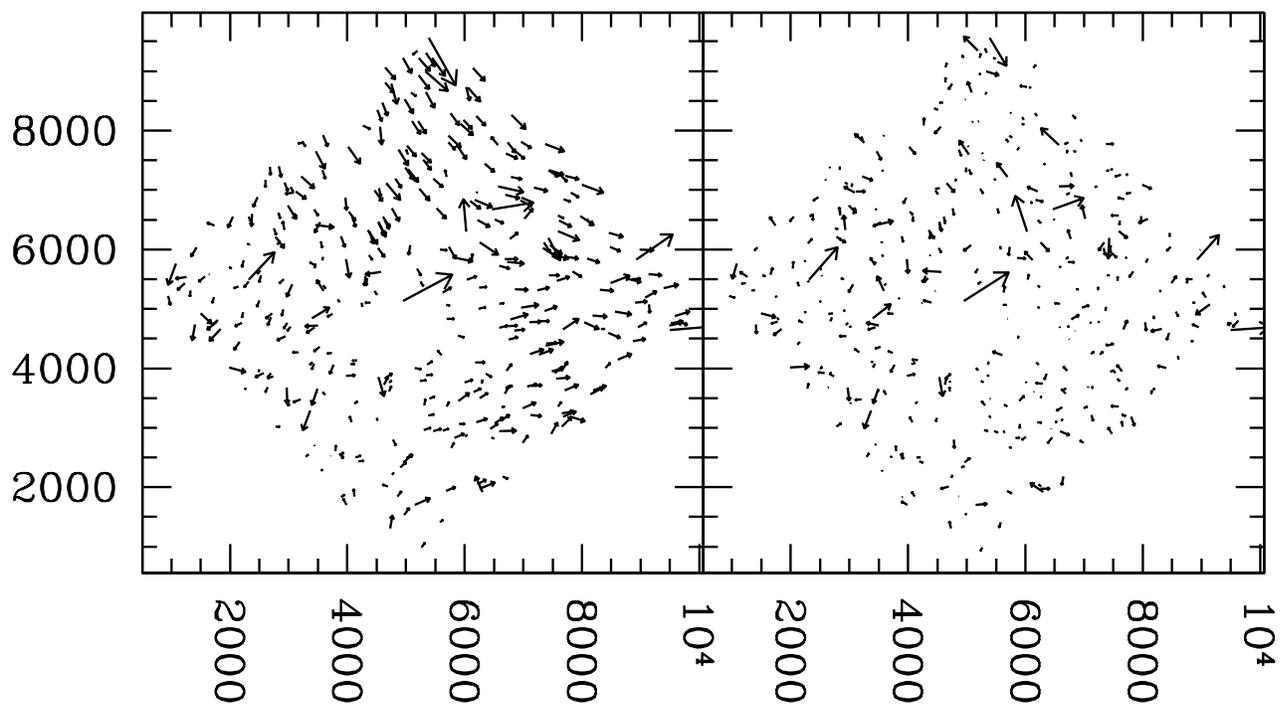}
\caption{Left panel: observed shifts (scaled by a factor of 500) of the \objpmnum most compact sources in the HUDF between Epoch 1 to Epoch 2, \mi\ band, images; Right panel:  corrected shifts, obtained after applying a correction based on a fit of the raw measurements (right panel, see Section \ref{PM}). North is up on these panels. The axes units are UDF pixels (0.030''/pixel).\label{figPM}}
\end{figure}

The Galactic halo is isotropic and has no or little overall rotation.  Halo objects are therefore expected to lag behind the rapidly rotating Galactic disk component (and  the local standard of rest) with an average velocity of $\approx 200 \pm 100 {\rm km/s\ }$. As done in \citet{oppenheimer2001}, it is assumed here that Galactic halo objects have a velocity  that is more than  $2 \sigma$  above the velocity expected for an object that is in the Galactic disk  \citep[i.e. $ \geqslant 100 {\rm \ km/s,}$][]{chiba2000}.\\
The HUDF image stacks (Section \ref{thedata}) were subtracted from one another and the results were examined to see if any object, whether classified to be extended or not based on its ${\sindex}_{\mz}$ value, had moved significantly (e.g. more than 2.5 UDF pixel) from one image to the next. No such object was found.\\
The motion of objects in the HUDF  was further investigated by carefully measuring the position of objects separately in each of the eight available \mi\ band HUDF partial image stacks. These positions were measured using the IDL task MPFIT2DPEAK \footnotemark, using a fixed width (FWHM=2.5 pixel) Gaussian fit. 
Initially, the positions of the \objpmnum most compact sources (${\sindex} \leqslant 0.1$ and $\mi \leqslant 28.5$ mag)  were examined. An average position and a standard deviation of the mean were computed for each object in the field using Epoch 1 and Epoch 2 HUDF image stacks separately. Average Epoch 1 and Epoch 2 positions were subtracted from one another to compute the observed shift of each object. These are shown (scaled up by a factor of 500) in the left panel of Figure \ref{figPM}.  Repeated measurements of the position of sources in the field  using HUDF image stacks taken within the same Epoch were always within 0.1 HUDF pixel, or 0.003''. Still, and as shown in Figure \ref{figPM}, there is a systematic, field dependent  disagreement between the Epoch 1 and Epoch 2 measurements. This is likely caused by Epoch 1 and Epoch 2 images having been taken at position angles that are 90\degr\ apart, and/or the effect of  telescope breathing, and/or very small amount of residual image distortion that is not taken into account in the current model of the ACS distortion maps.  To produce a  $3\sigma$ detection with these data, an object must have a measured proper motion larger than 0.54  HUDF pixel or $\mu \geqslant 0.081 {\rm \ ''/year}$. No object in Table \ref{table1} was observed to have  such a large proper motion.\\
The situation was however improved using a simple third order two dimensional polynomial fit to the observed distribution shown in the left panel of Figure \ref{figPM}. This fit, when applied as a correction,  results in significantly lower systematics across the field, as shown in  the right panel of Figure \ref{figPM}. The accuracy of the corrected shifts (of compact objects) between Epoch 1 and Epoch 2 is improved by a factor of 3, with a $3\sigma$ detection now corresponding to a shift of 0.18 HUDF pixel or $\mu \geqslant 0.027 {\rm \ ''/year}$. \\
The search for high proper motion objects in the HUDF was extended to all \tobjnum  objects in the field  by measuring and correcting (using the correction determined above) the observed shifts of these objects. Only four sources were found to have moved significantly (but with a significance slightly lower than $3\sigma$) from Epoch 1 to Epoch 2. These objects are UID 443, 9020, 911, and 7525 and they are all unresolved objects that are listed in Table \ref{table1}. Their measured shifts  are  0.17, 0.17, 0.14, and 0.15 HUDF pixels respectively, or approximately 0.02''/year,  which is smaller than our imposed $3\sigma$ limit. 
\footnotetext{\url{http://astrog.physics.wisc.edu/~craigm/idl/}}


\subsection{Extragalactic Unresolved Sources}\label{QSO}
Two unresolved, extra-galactic sources  with  $\mi \leqslant 27.0$ were identified (UID 6732, 9397). The GRAPES spectrum of these objects show prominent emission lines and the redshift of these objects is estimated to be z=3.2 and z=3.0, respectively \citep{chun2004}.  Identification of such objects using GRAPES spectra could easily be carried out down to magnitudes fainter than \mi=27.0 mag,  as demonstrated by the fact that several fainter extragalactic objects with $27.0 \leqslant \mi \leqslant 29.5$ were easily identified (e.g. UID 4120, 8157). Without spectral information, these objects would have  been misidentified as stars (e.g. UID 9397 would be a good candidate for a cool white dwarf at a distance of $\approx$ 1 Kpc). The issue of contamination by faint, blue extra-galactic objects in previous studies \citep{ibata1999,mendez2000}  has been discussed in the past \citep{kilic2004} and Figures \ref{VBV} and \ref{IVI} illustrate how these objects could be mistaken for a variety of stellar objects ranging from  M and L dwarfs to white dwarfs. Spectroscopic confirmation is therefore crucial to accurately determine stellar counts in deep fields such as the HUDF.  
There are \goodqsonum extra-galactic objects amongst the \psnumspc point sources with $\mi \leqslant 27.0$ mag, which  corresponds to a 7\% contamination level at these Galactic coordinates. \\
The \goodqsonum extra-galactic objects (UID 6732, 9397) were used to check the accuracy of the proper motion measurements described in Section \ref{PM}. The measured shifts of these objects between Epoch 1 and Epoch 2 are smaller than 0.04 HUDF pixel (0.0012''), which is what one would expect based on the proper motion error estimate quoted in Section \ref{PM}).\\
There are a total of 14 faint ($27.0 \leqslant \mi \leqslant 29.5$), blue ($(\mb-\mv) \leqslant  1.8$, $(\mv-\mi) \leqslant 0.8$) objects listed in Table \ref{table1} that were not included in the analysis above because they are either too faint to have useable GRAPES spectra or  robust \sindex based classifications. Some of these sources might well be faint white dwarfs in either the Galactic disk or halo but are hard to distinguish from extra-galactic sources without higher signal-to-noise spectra or more accurate proper motion measurements.

\section{Discussion}\label{discussion}
\subsection{M and L Dwarfs and the Galactic disk scale height}\label{mdwarfs}
\begin{figure}
\includegraphics[width=5.0in]{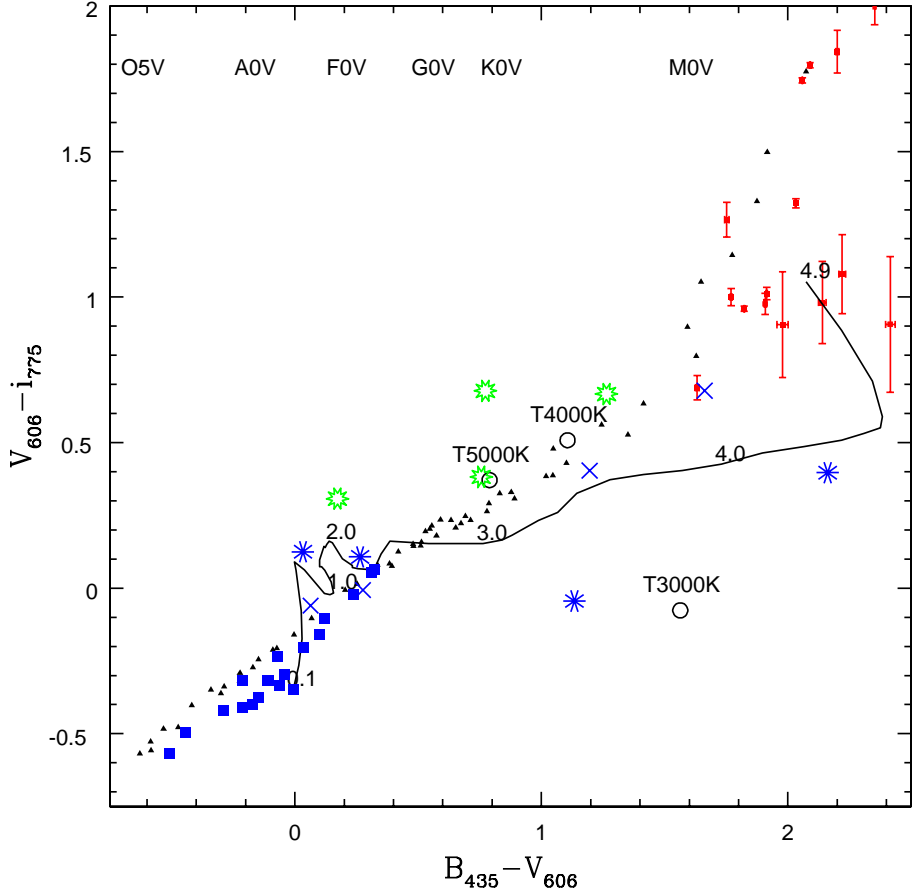}
\caption{(\mv-\mi) vs. (\mb-\mv) color-color plot of $\mi \leqslant 27.0$, unresolved, spectroscopically identified  objects in the HUDF. M and L dwarfs are shown using red squares with photometric error bars. Main-sequence stars are shown in green thick circles. Extra-galactic objects, including those dimmer than  $\mi=27.0$ mag, are shown using large blue stars. The 5 white dwarf candidates are shown using large blue crosses. Pickles main-sequence objects are also shown using small black circles. The (\mb-\mv) locations of the O, A , F, G, K, and M stellar type are shown for reference. Hot white dwarfs are shown in large blue squares. Cool white dwarfs   \citep[(3000\degr K, 4000\degr K, 5000\degr K),][]{harris2001} are shown using labeled, empty circles. Finally, the solid black line shows the QSO track, from redshift of z=4.0 (top right) to z=0.1 (bottom left)\label{vibv}}
\end{figure}

\begin{figure}
\includegraphics[width=5.0in]{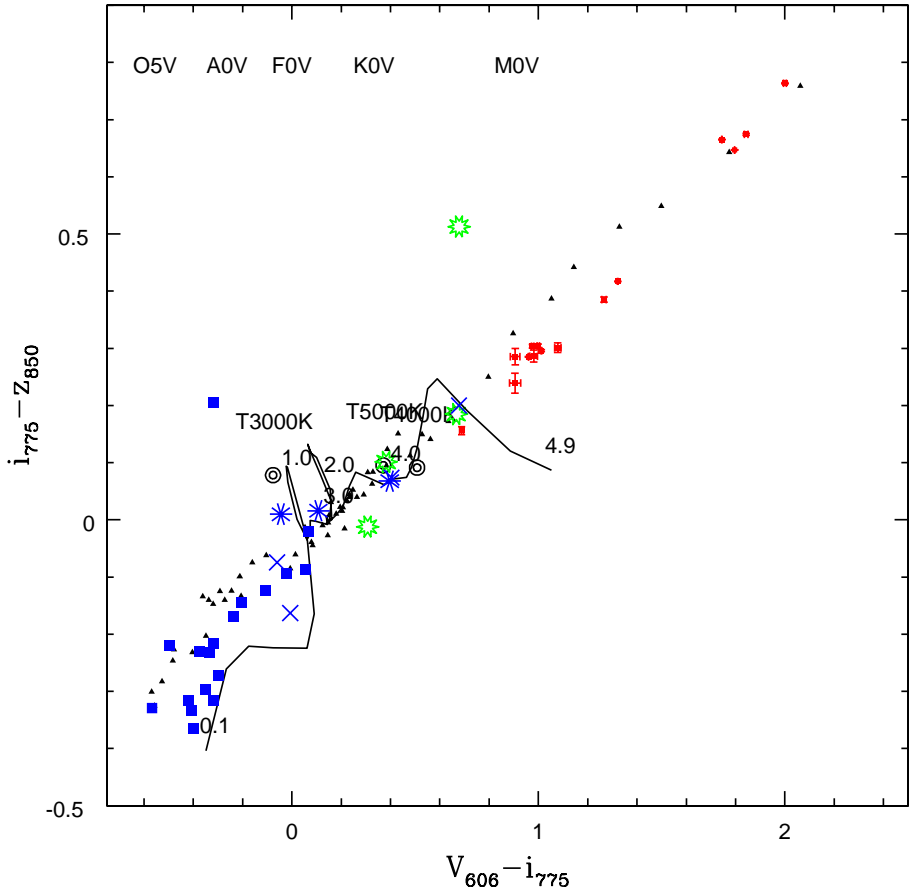}
\caption{(\mi-\mz) vs. (\mv-\mi) color-color plot of $\mi \leqslant 27.0$ mag, unresolved, spectroscopically identified  objects in the HUDF. The labels and symbols used are the same as in Figure \ref{vibv}.\label{izvi}}
\end{figure}

{\mdwawfsnum}  objects were identified to be M (or early L) dwarf main-sequence stars. Two objects were found to be best fit by early L dwarf templates \citep{kirkpatrick2000}. Both of these objects are red enough to be outside of the color range shown in Figures  \ref{vibv} and \ref{izvi} with $(\mv-\mi)=2.21$ and $(\mv-\mi)=3.47$. Deep infrared observations of these two objects will be required to confirm that they are L dwarfs. NICMOS observations of the HUDF did not detect these objects in either the J or H bands. All other very red objects are best fit using M dwarf templates. The broad spectral features of M dwarfs listed in Table \ref{table2} were well fitted using M dwarf \citet{pickles1998} templates, and for objects brighter than $\mi=27.0$ mag we believe the identification to be secure.\\
Because these late-type sources can be seen to great distances, they can be used to probe the structure of the Galactic disk and halo as traced by the lowest-mass stellar components of these populations. This was  explored  using Monte Carlo mass function (MF) simulations based on those developed by \citet{burgasser2004}.  Assuming power-law representations of the MF, dN/dM $\propto$ M$^{-\alpha}$, for masses 0.005 to 0.2 M$_{\sun}$, luminosity (LF) and effective temperature (T$_{eff}$) distributions were created using evolutionary models from \citet{burrows1997} over the range 100 $\lesssim$ T$_{eff}$ $\lesssim$ 4000 K.  For all simulations, $\alpha$ was fixed at 1.13 for $0.1 \lesssim {\rm M} \lesssim 0.2 {\rm \ M_{\sun}}$ \citep{reid1999} but allowed to vary between 0.5--1.5 for ${\rm M} \lesssim 0.1 {\rm \ M_{\sun}}$; number densities are normalized to the empirical value at 0.1 M$_{\sun}$ from the 8 pc sample \citep{reid1999}.

Two separate populations were simulated: a disk population with a flat age distribution spanning 10 Myr to 10 Gyr (the majority of sources have age
greater than 1 Gyr), and a halo population with uniformly sampled ages between 9-10 Gyr and a relative number density of 0.25\% \citep{digby2003}.
The resulting bolometric LFs were converted to $I$ (Cousins) LF's using a polynomial fit to empirical data (spanning $12.4 \lesssim {\rm M_I} \lesssim 22.2$; i.e., M4 to T8) from \citet{dahn2002} and \citet{golimowski2004}. Note that this fit is based on measurements for Solar metallicity field dwarfs, and may not be appropriate for a subsolar metallicity halo subdwarf population.
Apparent $I$-band distributions were determined by first assuming a constant density population out to a limiting magnitude m$_I$ = 27.5 (taking into
account the difference between HUDF AB-magnitudes and Vega magnitudes from the empirical data) and then applying a correction for the vertical distribution of sources. For the disk population, we assumed a density distribution $n \propto sech{(z/h_0)}^2$ \citep{reid2000}, where $z = d\sin{\beta}$, $d$ is the source distance, $\beta = 54.5\degr$ is the Galactic latitude of the HUDF, and $h_0$ is the 1/e disk scale height, assumed to range between 200 and 500 pc.  For the halo population, a Galactocentric oblate spheroid distribution as given in \cite{digby2003} was assumed, using the values derived there and an axial ratio q = c/a = 0.7.

\begin{figure}
\includegraphics[width=5.0in]{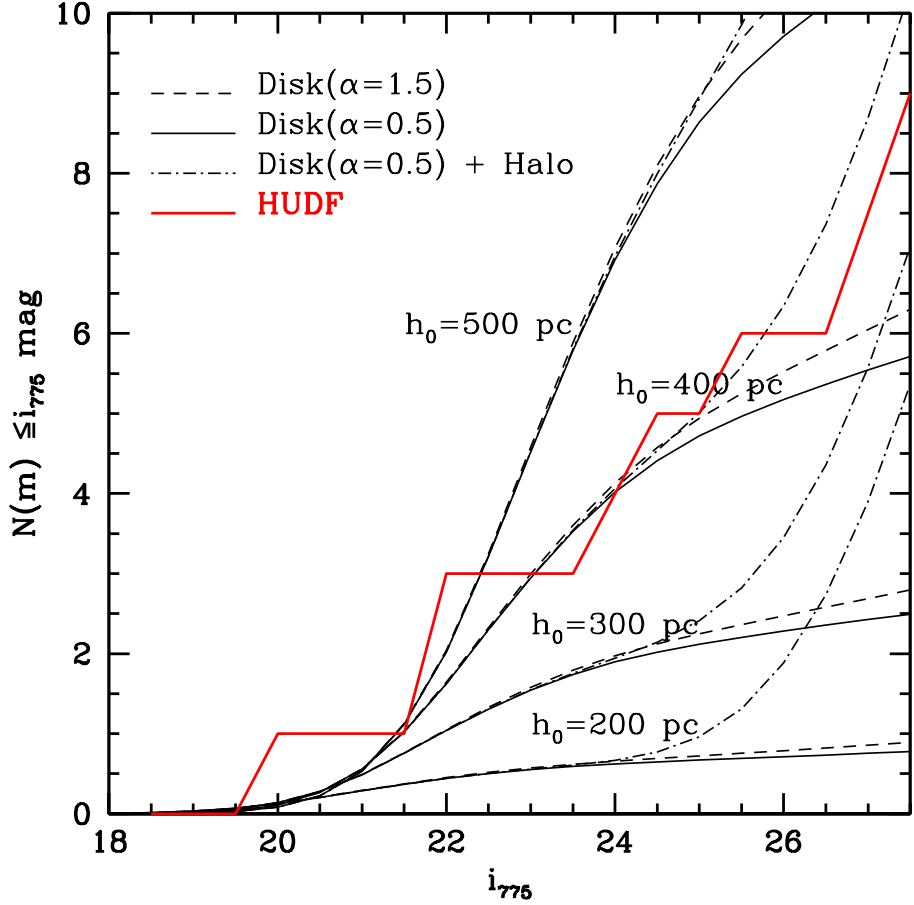}
\caption{Predicted and observed cumulative luminosity functions of stars with spectral types M4 and later. This plot shows the expected distribution from the Galactic disk and halo. Four different disk scale heights are plotted with $h_0$ values of 200, 300, 400 and 500 pc (bottom curves to top curves respectively). For each value of $h_0$, Mass Functions corresponding to $\alpha=0.5$ and 1.5 are shown (solid and dashed lines respectively). The observed HUDF distribution is shown using a solid black line which  is best fit by a disk with a scale height of $h_0=400 {\rm pc}$ while reasonably excluding values of $h_0$ = 200,300, and 500.\label{Mlf2}}
\end{figure}

Figure \ref{Mlf2}  presents the results of these simulations in the form of cumulative number distributions as a function of \mi\ magnitude down to the limiting magnitude of our sample. These distributions show that variations in the disk scale height are far more pronounced than those from the different MF's assumed, and this analysis is limited to the former parameter.  The cumulative distribution of the six dwarfs in Table \ref{table2} with spectral types M4 and later (consistent with the mass constraints of the simulations) is also shown.  The observed distribution matches that of the $h_0 = 400 {\rm \ pc}$ disk simulation very well, particularly out to $\mi \lesssim 23.5$ mag where the halo population makes negligible contribution.  Because the simulated distributions are rather sensitive to the disk scale height, particularly at fainter apparent magnitudes, one can conservatively constrain $h_0$ to $\pm$100 pc assuming Poisson uncertainties.  This is in good agreement with the estimate of $340 \pm 84 {\rm pc}$ from \cite{ryan2004} which was computed using ACS HUDF parallel fields i-drops.
For $\mi \geqslant 25$ mag, the observed distribution is slightly greater than the disk population alone but does not increase as sharply as the combined disk+halo distribution.  Indeed, the shape of the observed number distribution implies few if any halo stars in the HUDF down to $\mi = 27.0$ mag, consistent with the lack of significant proper motion sources in this sample.  This suggests that either the number
density ratio of halo to disk stars, or the adopted axial ratio for the halo density distribution, or both, may be smaller than assumed here. With no late-type halo subdwarf detections in this sample, we cannot usefully constrain these possibilities.

\subsection{White Dwarfs}\label{WD}
\subsubsection{White Dwarf candidates}
Objects brighter than  $\mi = 27.0$ were individually examined and distances were estimated under various assumptions of what the exact nature of each object might be:   (1) a main sequence star; (2) a young white dwarf; (3) an old white dwarf. The reddest objects, which are all well-fitted by M-and-later type templates, as well as the few extra-galactic objects identified in Section \ref{QSO}  were excluded from this analysis. The remaining \wdnum objects were all initially considered to potentially be white dwarfs.\\

Assuming that an object is a white dwarf, one can use the measured $\mv-\mi$ color of that object (Table \ref{table1}) and the cooling curves of \citet{richer2000} (after accounting for passbands and zeropoints differences), to derive an absolute \mv\ band magnitude, an age, and a distance to that object.  As shown in figure 2 of \citet{richer2000}, blue white dwarfs can either be young, hot objects (${\rm Age} \leqslant 5 {\rm \ Gyr}$, T $\approx 10^4$K, $M_{V_{606}} \leqslant  16.0$ mag) or  older, cooler, intrinsically dimmer ones (Age $\geqslant $ 10 Gyr, T $\approx 3 \times 10^3 {\rm \ K}$, $M_{V_{606}} \geqslant  16.0$ mag). This results in several white dwarf distance estimates for each object.\\
Proceeding via a process of elimination, the distinction between main sequence stars, young white dwarfs, old white dwarfs, and disk or halo white dwarfs is possible. Assuming that a star is a main sequence star, which is intrinsically much more luminous  than a white dwarf,  can cause the distance estimate for that object to be unreasonably large. Assuming that a star is a young white dwarf (${\rm Age} \leqslant  5.0 {\rm \ Gyr}$) implies that this object is less likely  to be  part of the Galactic halo since the latter is composed of  much older objects. Also, and based on the lack of proper motion detection discussed in Section \ref{PMm}, one can define an upper limit to  the tangential velocity (${V_T}_{max}$) of each object once its distance is determined. The motion of an object moving by more than 0.18 HUDF pixel (0.027 ''/ yr) should have been detected at the $3\sigma$ significance level as discussed in Section \ref{PM}. As discussed above, objects in the Galactic halo are expected to have velocities around 100-200 km/s.  Even if  projection effects should result in lower values of   ${V_T}$, one would expect at least some of them to have  ${V_T}$ values larger than 30-60 km/s. A final clue to help narrow down the nature of a particular object is provided by the fact that an object that is nearby would suffer from a significant, easily measurable, parallax during the 73 days interval between Epoch 1 and Epoch 2 observations. Calculating the parallax vector for the HUDF exposures shows that an object 30 pc away would produce a parallax of about 0.060'' or 2 HUDF pixel between Epoch 1 and Epoch 2 observations. Objects that are closer than 200 pc are therefore expected to produce a parallax that would be detected at more than $3\sigma$ level. \\

Examining Table \ref{table4}, objects 4322, 4839, 7768, 9020 are unlikely to be main-sequence stars as this would make them very distant objects. The  Magellanic Stream  \citep{mathewson1984} is too far away from the HUDF to possibly explain the existence of main-sequence stars at such large distances. Objects 1147, 3166, and 5921 which do not (this is true of all objects listed in Table \ref{table4}) have measurable parallax or proper motion, are not likely to be white dwarfs. Being either young or old white dwarfs would  place these two objects less than 200 pc away and would have  resulted in measurable parallax in the HUDF image stacks. Another object unlikely to be a white dwarf is object 9230 which was observed to have a fainter, unresolved companion 0.5'' away. This companion, for which we have no GRAPES spectrum,  has a \mv\ band magnitude of $\mv=23.7$ mag and the colors  $\mb-\mv=1.6, \mv-\mi=0.9, \mi-\mz=0.4$ mag. Based on these colors, this object should be an early M dwarf that is $(4.9 \pm 1.2) \times 10^{3} {\rm \ pc}$ away. This places both object 9230 and its companion at the same distance (within error bars). The likelihood of any two stars being within 0.5'' in the HUDF, and at nearly the same distance, is very small ($\leqslant 5 \times 10^{-3}$) and these two objects are likely to be part of a binary system, with object 9230 being a K type main-sequence star. Independently of this, object 9230 cannot be an old white dwarf since this would place this object at distance of 34 pc where its parallax motion would have been very easily detected.\\

\subsubsection{Disk or halo white dwarfs?}\label{disk}
Four objects (4322, 4839, 7768, 9020) remain as white dwarf candidates. Using a simple 1/$V_{max}$ \citep{schmidt1968,tinney1993,mendez2002} analysis,  the white dwarf number density in the direction of the HUDF can be computed. The detection of 4 white dwarfs in the HUDF implies a local density of  $(3.5 \pm 1.5) \times 10^{-5}  \leqslant {\rm \ stars/pc^3} \leqslant (1.1 \pm 0.3) \times 10^{-2}$. The upper and lower limits on the density are computed using old or young white dwarfs respectively, while  the errors reflect the large uncertainties in the intrinsic luminosities of young and old white dwarfs ($\sim1.0$ mag). Note that this estimate is actually not affected by whether or not object 9230 was included as a HUDF white dwarf.  The contribution of this object to the 1/$V_{max}$ is negligible.  For comparison, \citet{liebert1988} previously determined the local white dwarf density to be $3.0 \times 10^{-3} {\rm \ stars/pc^3}$, while \citet{reid2001} found a value of $(3.26 \pm 1.23) \times 10^{-3} {\rm \ stars/pc^3}$, and  \citet{mendez2002} derived a density of white dwarfs in the thick disk of $(2.610 \pm 0.59) \times 10^{-2} {\rm \ stars/pc^3}$. The white dwarf density derived here is consistent with most of the previous density estimates. Properly determining the disk and halo membership of the 4 white dwarfs identified in the HUDF will be needed before a stronger conclusion can be made. Re-observing the HUDF field in the \mi\ band in 1.5 years would allow (assuming that a  0.18 HUDF pixel shift corresponds to a $3\sigma$ level detection) one to unambiguously detect the proper motion of halo white dwarfs with  ${V_T} \geqslant 100$ km/s at distances up to 1000 pc.\\
The above density estimate assumed that all four HUDF white dwarfs were Galactic disk objects. Could some of these objects be in the Galactic halo? At least two of these objects (4839, 7768) have $V_{Tmax}$ values that are lower than 60 km/s. The probability that this is caused by a projection effect  is under 10\% (assuming a relatively low velocity of  $\approx100$ km/s for halo objects). The last two white dwarfs listed in Table \ref{table4} (4322, 9020) are at  slightly larger distances which average to 583 pc, and more importantly have larger values of $V_{Tmax}$. The average distance of these objects still seems a bit low for objects which would be part of a Galactic halo whose density increases all the way up to 7 kpc  \citep{reid1993}, while white dwarfs with absolute magnitudes of $M_{V_{606}}=17$ mag \citep{harris2003} should be detected all the way out to 1000 pc. Unless these objects are intrinsically much dimmer, one would expect the average distance to these two objects to be larger than 583 pc and much closer to our limit of 1000pc. The $V_{Tmax}$ values inferred for objects 4322 and  9020, while higher than those of objects 4839 and 7768, fall substantially short of expected typical halo velocities. The probability of observing each of the objects with the  $V_{Tmax}$  values  listed in Table \ref{table4} is between 14--20\%. As a group, the  probability that objects 4322 and 9020 are halo objects is under 3\%.  Similarly, the hypothesis that all 4 detected white dwarfs in the HUDF are halo rather than disk objects is excluded at the 99.9986 \% level. Overall, the distances to the four white dwarf candidates identified here is more consistent with them being part of a Galactic thick disk. \citet{majewski2002} derived a disk scale height of 400--600 pc using a study of low velocity white dwarfs. This value was somewhat larger than previous estimates of 250--350 pc but it is interesting to note that the M-dwarf disk scale height derived in Section \ref{mdwarfs}, as well as the distance estimates of the white dwarfs listed in Table \ref{table4} all appear to agree with this value.\\
It would be interesting to attempt to set an upper limit on the halo white dwarf density based on the finding of up to two white dwarfs  in the HUDF. Under this assumption, the upper limit to the Galactic halo white dwarf density is computed to be  $(6.53 \pm 3) \times 10^{-3} {\rm \ stars/pc^3}$.  This result is consistent with the previous work of \citet{mendez2000} using the combined observations of the Hubble Deep Fields North and South ($7.73 \times 10^{-3} {\rm stars/pc^3}$). 
\\
A population II halo white dwarf population that is older than 12-13 Gyr would have remained undetected here (An intrinsically fainter than $M_{V_{606}}=17.0$ mag white dwarf population  would also help explaining the low average distances of the white dwarfs listed in Table \ref{table4}). As discussed above, the white dwarfs identified in this work are more likely to be within the Galactic thick disk. It is to be noted that  \citet{kilic2004} recently observed a disproportionately large fraction of disk to halo white dwarfs in the Hubble Deep Field North (HDF-N).  \citet{kilic2004} found no evidence of any white dwarfs with tangential velocities larger than 30 km/s down to $\mi=27.5$ mag and the authors concluded that the blue objects they saw were all part of the Galactic disk. Similarly, the HUDF observations presented here successfully identified 4 white dwarf candidates with $\mi \leq 27.0$ mag  in the HUDF,   none very likely to be in the Galactic halo while 2--3 detection would have been expected to be consistent with previous studies. \\ 

\subsubsection{Dark halo white dwarfs?}
While dynamical studies of the Galaxy predict the dark matter white dwarf density as high as  $1.26 \times 10^{-2} {\rm \ stars/pc^3}$ (assuming that 100\% of the dark matter halo mass is in the form of  $0.6 M\odot$ white dwarfs), the failure to detect a significant population of high velocity white dwarfs in the HUDF  points to a dark matter halo devoid of a significant white dwarf population. 
Following the methodology of \citet{kawaler1996}, one can estimate the amount of dark matter halo mass that is probed by the HUDF images, and directly compute the fraction of the dark matter halo mass that could be contained in a white dwarf population. Assuming a white dwarf absolute ${\rm M_{\mi}}$ magnitude of 17.0 mag\citep{harris2003} and a limiting magnitude of $\mi=27.0$ mag, the maximum probed distance is 1000 pc. In the corresponding volume, and in the direction of the HUDF,  the included dark matter halo mass is  3 $M_{\odot}$.
A population of dark matter halo white dwarfs could possibly have remained undetected in this study for several reasons:  (1) The method used in this paper may be inefficient at  properly identifying halo white dwarfs in Table \ref{table4}. In this case, if one assumes that the HUDF contains 4 halo white dwarfs ($\mi \leqslant 27.0$), these would  account for $2 M_{\odot} $, a significant fraction of the expected dark matter halo mass. This would however increase the ratio of halo to disk white dwarfs to unrealistically high levels (i.e. $\gg 2 \%$), and would assume that all the white dwarfs seen in the HUDF are from the dark matter halo and not from the pop II halo, which is highly unrealistic. The lack of proper motion detections also makes this scenario unlikely; (2) White dwarfs do not contribute significantly to the Galactic dark matter halo mass.  From microlensing experiements towards the Magellanic clouds, the contribution of MACHOs to dark matter has been estimated to be anywhere between 20\%  to less than 2\% \cite[][and references therein]{alcock2000, afonso2003, sahu2003, evans2004}. Furthermore, \cite{majewski2002} showed how an increased thick disk scale height of 400-600 pc makes it unlikely that white dwarfs could be the MACHO objects; (3) White dwarfs  could contribute significantly to the mass of the Galactic dark matter halo but are too faint to be detected in this study. \citet{richer2000} showed that  an increase in the age of the white dwarf halo population from 10 Gyr to 16 Gyr is expected to reduce the number of white dwarf detection (down to $\mv = 28.0$) by a factor of 7.
The lack of high velocity white dwarf detections in the HUDF puts upper limits on the contribution of an hypothetical white dwarf population to the Galactic dark matter halo. Based on the brightest objects in the HUDF for which we have spectra (i.e. $\mi \leqslant 27.0$), if the dark matter halo is as old as about 12 Gyr \citep{hansen2002,charboyer1998} then dark matter halo white dwarfs with $M_V=17.5$ ($M_\mi= 17.0$) contribute less than 20\% to the dark matter. If the age of the dark matter halo is  10 Gyr, white dwarfs ($M_\mi= 15.9$) contribute less than 2\% to the dark matter.

The possibility that there is a large population of faint white dwarfs that has remained unidentified in this study and which would account for a significant fraction of the dark matter halo can be investigated a little further. Recall that in Section \ref{QSO}, 14 sources with ($27.0 \leqslant \mi \leqslant 29.5$) and blue ($(\mb-\mv) \leqslant  1.8$, $(\mv-\mi) \leqslant 0.8$) colors were identified. Not all of these sources were expected to be bona-fide stars and some are likely to be extra-galactic sources, but as discussed in  Section \ref{test}, it is estimated that 95\% of the stars in the HUDF were properly identified, so that this number should be considered to be an upper limit on the number of faint starts in the HUDF. Following the same methodology used above, but now using a limiting magnitude of  $\mi \leqslant 29.5$, the HUDF images probe through 79 $M_{\odot}$ of the dark matter halo. Even under the extreme assumption that all 14 unresolved candidates in the UDF are high velocities white dwarfs, this implies that, when reaching down to \mi=29.5, faint white dwarfs in the Galactic dark matter halo contributing less than 10\% to the total dark matter halo mass for a dark matter halo age of 12 Gyr. Future observations of the HUDF would allow more sensitive  measurements of the proper motion of these faint objects (Section \ref{disk}). Setting tighter constraints on the maximum tangential velocities of these faint HUDF unresolved objects would allow to exclude some of these objects from  high velocity white dwarfs and would allow to further lower the maximum fraction of the dark matter halo mass that could be accounted for by white dwarfs.

\section{Conclusions}
A systematic search for unresolved objects in the HUDF identified \psnum objects with $\mi \leqslant 29.5$ mag.  Using the GRAPES spectra of these objects, \psnumspc objects with $\mi \leqslant 27.0$ mag were spectroscopically identified, including \mdwawfsnum M and later dwarfs (including  2 unconfirmed L dwarf candidates), and \goodqsonum QSOs. The M dwarf luminosity function of M4 and later-type stars was computed and compared to predictions based on Monte-Carlo simulations. Assuming a simple analytical model of the Galactic disk, the number of detected M dwarfs was shown to be large enough to constraint the scale height of the M and L dwarf Galactic disk to be $h_0=400 \pm 100 {\rm pc}$.
Out of 8 stars that were found to not be M or later-type stars, four stars were determined to be old white dwarf candidates. Not a single object was found to have a proper motion that is larger than 0.027''/year making them likely Galactic thick disk objects. Further imaging of the HUDF, using a time span larger than 1.5 year would provide more sensitive proper motion measurements. More sensitive proper measurements are required to positively place unresolved objects in the Galactic disk or halo. It would also allow to search for high tangential velocity objects amongst the 14 fainter ($27.0 \leqslant \mi \leqslant 29.5$) unresolved sources identified in the HUDF. Excluding halo membership of objects down to $\mi=29.5$ based on proper motion measurements would  further constraint the maximum white dwarf contribution to the dark matter halo mass. The currently available observations of the HUDF, spanning 73 days, show the absence of a significant population of high velocity white dwarfs down to $\mi = 27.0$, and a relatively small number of unresolved, faint blue sources in the field down to $\mi=29.5$. This is interpreted as a consequence of  white dwarfs accounting for less than 10\% of the dark matter halo, assuming that the dark matter halo is 12 Gyr old. 

\acknowledgments
We would like to thank I. N. Reid for helpful discussions. This work was supported by grant GO -09793.01-A from the Space Telescope Science Institute, which is operated by AURA under NASA contract NAS5-26555.

\begin{deluxetable}{cccccccccc}
\tabletypesize{\footnotesize} 
\tablecaption{Unresolved objects in the HUDF, as selected using the criterion ${\sindex}_{\mz} \leqslant $0.05 (F850LP ACS band). \label{table1}}
\tablehead{\colhead{UID} & \colhead{RA (J2000)} & \colhead{Dec (J2000)} & \colhead{\mb} & \colhead{\mb-\mv}   &  \colhead{\mv-\mi} & \colhead{\mi-\mz} & \colhead{${\sindex}_\mi$} & \colhead{${\sindex}_\mz$}
}
\startdata  
19 &  53.1623248 &  -27.8269460 & 26.58  & 1.91 & 1.01 & 0.30 & 0.015 & 0.004\\
366 &  53.1753062 & -27.8198904  & 29.99   & 2.20 & 3.05 & 1.27 & 0.009 & 0.030\\
443 &  53.1583988 & -27.8189953  & 31.25  & 0.86 & 3.47 & 1.51 & 0.016 & 0.028\\
834 &  53.1648156 & -27.8143652  & 27.67   & 2.35 & 2.00 & 0.76 & 0.013 & 0.011\\
911 &  53.1460739 & -27.8123237  & 25.21   & 2.09 & 1.80 & 0.65 & 0.007 & 0.002\\
1147 & 53.1834639 & -27.8067589  & 19.58  & -0.16 & 0.52 & 0.13 & 0.091 & 0.002 \\
1343 &  53.1398080 & -27.8100295  & 27.36   & -0.23 & -0.14 & -0.04 & 0.045 & 0.046\\
2150 & 53.1767483 & -27.7996684  & 22.60   & 1.99 & 1.43 & 0.80 & 0.147 & 0.003 \\
2368 &  53.1676057 & -27.8037923  & 29.07   & 1.18 & 0.08 & -0.20 & 0.046 & 0.050\\
2457 &  53.1616105 & -27.8027738  & 26.98  & 1.91 & 0.98 & 0.30 & 0.004 & 0.008\\
2977 &  53.1726388 & -27.8006672  & 27.94   & -0.18 & -0.21 & 0.18 & 0.042 & 0.049\\
3166 & 53.1583425 & -27.7949215  & 20.80   & 0.96 & 0.80 & 0.37 & 0.086 & 0.007 \\
3561 &  53.1483050 & -27.7977856  & 29.24  & 1.98 & 0.91 & 0.24 & 0.018 & 0.012\\
3794 &  53.1474097 & -27.7965561  & 28.89   & 2.14 & 0.98 & 0.29 & 0.010 & 0.007\\
3940 &  53.1498728 & -27.7961691 &  27.06   & -0.48 & -0.21 & -0.18 & 0.033 & 0.038\\
4120 &  53.1839107 & -27.7954147  & 28.04   & 0.03 & 0.12 & 1.31 & 0.023 & 0.012\\
4322 & 53.1349058 &  -27.7944480  & 26.83   & 0.06 & -0.06 & -0.07 & 0.067 & 0.043\\
4643 & 53.1848922 & -27.7933971   & 31.73   & 1.52 & 0.74 & 0.25 & 0.019 & 0.050\\
4839 &  53.1883621 &   -27.7923200  & 27.22  & 1.20 & 0.40 & 0.07 & 0.018 & 0.015\\
4945 &  53.2003709 & -27.7898451 &  25.06  & 2.06 & 1.74 & 0.67 & 0.020 & 0.013\\
5317 &  53.1257144 & -27.7904647  & 29.06  & 0.49 & -0.00 & -0.30 & 0.045 & 0.039\\
5441 &  53.1625253 & -27.7896458  & 27.81   & 1.75 & 1.27 & 0.39 & 0.016 & 0.019\\
5921 & 53.1322881 & -27.7828516 & 20.57 & 0.77 & 1.19 & 0.51 & 0.094 & 0.070\\
5992 &  53.1498432 & -27.7874448 &  27.52   & 1.63 & 0.69 & 0.16 & 0.014 & 0.005\\
6334 &  53.1782000 & -27.7861688  & 27.71   & -0.27 & -0.16 & -0.18 & 0.049 & 0.049\\
6442 &  53.1416300 & -27.7856726 &  28.16 & -0.19 & 0.04 & -0.07 & 0.044 & 0.030\\
6461 &  53.1984259 & -27.7848674  & 26.80  & 1.77 & 1.00 & 0.30 & 0.010 & 0.010\\
6620 &  53.1428439 & -27.7849376  & 31.27   & 3.19 & 0.25 & -0.14 & 0.041 & 0.049\\
6732 &  53.1784892 & -27.7840395  & 25.73  & 1.13 & -0.04 & 0.01 & 0.010 & 0.022\\
7113 &  53.1707114 & -27.7826303  & 30.83   & 2.49 & 0.65 & 0.20 & 0.026 & 0.024\\
7194 &  53.1860335 & -27.7822207  & 26.91  & -0.12 & -0.13 & -0.06 & 0.048 & 0.045\\
7357 &  53.1692789 & -27.7813943 &  27.93   & 0.10 & -0.21 & -0.16 & 0.024 & 0.031\\
7525 &  53.1318579 & -27.7820076 &  28.70   & 2.22 & 1.08 & 0.30 & 0.011 & 0.007\\
7768 &  53.1472186 & -27.7714786  & 27.12   & 1.66 & 0.68 & 0.20 & 0.014 & 0.005\\
7894 &  53.1831085 & -27.7802451  & 27.46   & 0.03 & -0.15 & -0.12 & 0.047 & 0.039\\
8081 &  53.1580838 & -27.7701503  & 27.92   & -0.32 & -0.40 & -0.30 & 0.038 & 0.020\\
8157 &  53.1649526 & -27.7736713  & 30.85   & 2.16 & 0.40 & 0.07 & 0.042 & 0.039\\
8186 &  53.1867094 &  -27.7735490 &  27.72   & -0.13 & -0.14 & -0.34 & 0.044 & 0.029\\
9006 &  53.1853179 & -27.7799976  & 27.46   & -0.05 & -0.07 & 0.24 & 0.039 & 0.024\\
9020 &  53.1685421 & -27.7805214 &  27.00   & 0.28 & -0.01 & -0.16 & 0.025 & 0.034\\
9212 &  53.1485421 & -27.7701387  & 25.50   & 1.82 & 0.96 & 0.29 & 0.011 & 0.003\\
9230 &  53.1580164 & -27.7691869  & 21.18  & 0.76 & 0.38 & 0.10 & 0.037 & 0.007\\
9331 &  53.1638484 &  -27.767128  & 27.95   & 2.20 & 1.84 & 0.67 & 0.025 & 0.005\\
9351 &  53.1781506 & -27.7691241  & 26.21   & 2.03 & 1.32 & 0.42 & 0.015 & 0.004\\
9397 &  53.162852 & -27.7671662  & 21.44   & 0.26 & 0.11 & 0.02 & 0.043 & 0.033\\
9959 &  53.1611698 & -27.7555109  & 29.12   & 2.42 & 0.91 & 0.29 & 0.022 & 0.021\\
\enddata

\end{deluxetable}

\begin{deluxetable}{ccccccc}
\tablecaption{Spectral types of the \psnum  unresolved objects in the HUDF (\psnumspc brighter than $\mi=27.0$ mag) and their distance estimates assuming that they are main-sequence objects or white dwarfs with $M_V=14$ or $M_V=17$. Two additional, fainter extra-galactic sources are included (UID 4120, 8157) The spectral types were determined by fitting the GRAPES \citep{pirzkal2004} spectra of these sources to a set of templates (See Section \ref{stars}).\label{table2}}
\tabletypesize{\footnotesize} 
\tablehead{\colhead{UID} & \colhead{${\rm S_{type}}$} & \colhead{$\mi \geqslant 27.0$} &  \colhead{${\rm M_{V_{606}}}$} &    \multicolumn{3}{c}{Log(D)}  \\
& & & & MS & WD $M_V=14$ & WD $M_V=17$}
\startdata
19 & M1-M2 & no & $9.30\pm 0.25$ & $4.072\pm 0.050$ & 3.129 & 2.529 \\
366 & L0-L1 & no & $18.9\pm 0.50$ & $2.780\pm 0.100$& 3.750 & 3.150 \\
443 & L0-M6 & no & $15.6\pm 2.9$ & $ 4.261\pm 0.570 $&  4.274 & 3.674 \\
834 & M4-M5 & no & $11.4\pm 0.39 $ & $ 3.777\pm 0.077 $&  3.258 & 2.658 \\
911 & M4-M5 & no & $11.4\pm 0.39 $ & $ 3.338\pm 0.077 $&  2.818 & 2.218 \\
1147 & F6-F8 & no & $3.82\pm 0.17 $ & $ 4.180\pm 0.033 $&  2.142 & 1.542 \\
1343 & A0-K7 & yes & $4.32\pm 3.58 $ & $ 6.080\pm 0.716 $&  3.713 & 3.113 \\
2150 & M3-M4 & no & $11.3\pm 0.23 $ & $ 2.862\pm 0.045 $&  2.315 & 1.715 \\
2368 & O9-M6 & yes & $4.19\pm 8.55 $ & $ 7.145\pm 1.710 $&  3.774 & 3.174 \\
2457 & M0-M1 & no & $8.77\pm 0.28 $ & $ 4.258\pm 0.056 $&  3.209 & 2.609 \\
2977 & A0-M4 & yes & $11.28\pm 0.23 $ & $ 4.365\pm 0.045 $&  3.818 & 3.218 \\
3166 & K4-K5 & no & $6.96\pm 0.14 $ & $ 3.573\pm 0.027 $&  2.164 & 1.564 \\
3561 & M0-M1 & no & $8.77\pm 0.28 $ & $ 4.697\pm 0.056 $&  3.647 & 3.047 \\
3794 & M0-M1 & no & $8.77\pm 0.28 $ & $ 4.594\pm 0.056 $&  3.544 & 2.944 \\
3940 & O5-M3 & yes & $2.37\pm 7.94 $ & $ 7.315\pm 1.588 $&  3.702 & 3.102 \\
4120 & GAL & yes &     &   &      &   \\
4322 & A7-K5 & no & $4.70\pm 2.4 $ & $ 5.632\pm 0.479 $&  3.548 & 2.948 \\
4643 & A0-M6 & yes & $6.75\pm 5.96 $ & $ 6.578\pm 1.191 $&  4.237 & 3.637 \\
4839 & G5-K2 & no & $5.65\pm 0.61 $ & $ 5.087\pm 0.122 $&  3.400 & 2.800 \\
4945 & M4-M5 & no & $11.4\pm 0.38 $ & $ 3.315\pm 0.077 $&  2.796 & 2.196 \\
5317 & A0-M6 & yes & $6.71\pm 6.0 $ & $ 6.267\pm 1.199 $&  3.910 & 3.310 \\
5441 & M2-M3 & no & $9.93\pm 0.38 $ & $ 4.228\pm 0.076 $&  3.407 & 2.807 \\
5921 & K4-K5 & no & $6.96\pm 0.14 $ & $ 3.565\pm 0.027$&  2.155 & 1.555\\ 
5992 & K7-M0 & no & $8.07\pm 0.43 $ & $ 4.568\pm 0.085 $&  3.373 & 2.773 \\
6334 & F8-L0 & yes & $8.34\pm 4.4 $ & $ 5.505\pm 0.875 $&  3.791 & 3.191 \\
6442 & A0-G5 & yes & $2.89\pm 2.2 $ & $ 6.271\pm 0.429 $&  3.865 & 3.265 \\
6461 & M0-M1 & no & $8.77\pm 0.28 $ & $ 4.251\pm 0.056 $&  3.202 & 2.602 \\
6620 & B8-M4 & yes & $5.45\pm 5.6 $ & $ 6.345\pm 1.121 $&  3.812 & 3.212 \\
6732 & GAL & no &     &   &      &   \\
7113 & G8-K0 & yes & $5.60\pm 0.19 $ & $ 5.543\pm 0.038 $&  3.861 & 3.261 \\
7194 & O5-K5 & yes & $0.76\pm 6.3 $ & $ 7.215\pm 1.266 $&  3.601 & 3.001 \\
7357 & O9-M5 & yes &$3.73\pm 8.1 $ & $ 7.131\pm 1.618 $&  3.760 & 3.160 \\
7525 & M1-M2 & no & $9.33\pm 0.23 $ & $ 4.429\pm 0.045 $&  3.491 & 2.892 \\
7768 & K5-K7 & no & $7.37\pm 0.27 $ & $ 4.616\pm 0.055$&  3.286 & 2.686 \\
7894 & B8-K7 & yes & $3.73\pm 3.9 $ & $ 6.223\pm 0.780 $&  3.680 & 3.080 \\
8081 & O5-A3 & yes & $-1.98\pm 3.6 $ & $ 7.471\pm 0.716 $&  3.843 & 3.243 \\
8157 & GAL & yes &     &   &    &   \\
8186 & B0-K7 & yes & $1.89\pm 5.8 $ & $ 7.038\pm 1.151 $&  3.764 & 3.164 \\
9006 & B9-A3 & yes & $0.94\pm 0.65 $ & $ 6.328\pm 0.131 $&  3.697 & 3.097 \\
9020 & F0-F2 & no & $3.16\pm 0.44 $ & $ 5.716\pm 0.089 $&  3.540 & 2.940 \\
9212 & M0-M1 & no & $8.77\pm 0.28 $ & $ 3.979\pm 0.056$&  2.929 & 2.329 \\
9230 & K2-K3 & no & $6.40\pm 0.14 $ & $ 3.800\pm 0.028 $&  2.279 & 1.679 \\
9331 & M5-M6 & no & $12.3\pm 0.45 $ & $ 3.701\pm 0.089 $&  3.345 & 2.745 \\
9351 & M1-M2 & no & $9.30\pm 0.25 $ & $ 3.974\pm 0.050 $&  3.031 & 2.431 \\
9397 & GAL & no &     &   &   &   \\
9959 & M0-M1 & no & $8.77\pm 0.28 $ & $ 4.585\pm 0.056 $& 3.535 & 2.935 \\
\enddata
\end{deluxetable}

\begin{deluxetable}{cccccccc}
\rotate
\tablehead{\colhead{UID} & \colhead{Type} & \colhead{${\rm M_{\mv}}$} & \colhead{D} & \colhead{Age} & \colhead{${V_T}_{max}$} &   \colhead{Possible} & \colhead{1/$V_{max}$}\\
& & & \colhead{(pc)} & \colhead{(Gyr)} & \colhead{(km/s)} &   \colhead{Type}  & \colhead{($star/pc^{-3}$)} 
}
\tablecaption{White dwarf candidates in the HUDF ($\mi  \leqslant $27.0 mag). We list the derived distance to each object under the assumptions that it is a  main sequence star or  white dwarfs \citep{richer2000} with the appropriate $V-i$ colors (See Table \ref{table1}).  ${V_T}_{max}$ is the {\em upper limit} of the tangential velocity ($V_T$) of each object. The seventh column lists the acceptable nature of the object  while the last column gives the 1/$V_{max}$ density for that particular object/scenario.  Section \ref{WD} explains how these quantities were derived in more detail.\label{table4}}

\tabletypesize{\footnotesize} 

\startdata
1147 & MS F6-F8 & 3.7-4.0  & $1.5 \pm 1.1 \times 10^4$  & & & F6-F8 \\
1147 & Young WD  & 14.0-15.0 & 110 $\pm$ 24  & 1.4-3.9 & 14.1 & & \\
1147 & Old WD & 17.4-18.3 & 24 $ \pm$ 5 &  10.3-12.1 & 3.0 & & \\
 \hline
3166 & MS K4-K5 & 6.8-7.1& $3.7 \pm 0.23 \times 10^3$ & & & K4-K5\\
3166 & Young  WD & 14.8-15.8 & 80$ \pm$ 18 & 2.5-5.7 & 10.2 & & \\
3166 & Old WD & 17.2-18.2 & 27 $\pm$ 6 & 9.9-11.5 & 3.5 & & \\
 \hline
4322 & MS A7-K5&2.3-7.1 & $4.29 \pm 3.4  \times 10^5$ & &  \\
4322 & Young WD  & 12.1-13.1 & $6.9  \pm 1.6 \times 10^3$ & 0.4-1.4 & 1545.6 & \\
4322 & Old WD & 17.5-18.4 & 590 $\pm$ 126 & 10.7-12.9 &  75.5 & Disk or Halo & $3.37 \times 10^{-3}$\\
 \hline
4839 & MS G5-K2& 5.0-6.3 & $1.2 \pm 0.3 \times 10^5$ & &  \\
4839 & Young WD  & 12.5-13.5 & $2.4  \pm  0.5 \times 10^3$  & 1.1-3.3 & 885. & Disk & $9.45 \times 10^{-6}$\\
4839 & Old WD & 17.4-18.3 & 417 $\pm$ 88 & 10.4-12.3 &  53.4 & Disk or Halo & $1.80 \times 10^{-2}$\\
 \hline
 5921 & MS K4-K5 & 6.8-7.1 & $3.7 \pm 0.2 \times 10^3$ & & & K4-K5\\
 5921 & Young WD & 14.5-15.5 & $93\pm 20$ & 1.9-4.7 & 11.9 & &\\
 5921 & Old WD & 17.3-18.2 & $26\pm 5$& 10.1-11.7& 3.3 & & \\
 \hline
7768 & MS K5-K7& 7.1-7.6 & $4.1 \pm 0.5 \times 10^4$ & &  &\\
7768 & Young WD  & 14.5-15.5 & $1.3 \pm  0.3 \times 10^3$& 1.9-4.7 &   160.2 & Disk & $2.1 \times 10^{-5}$\\
7768 & Old WD & 17.3-18.2 & 345 $\pm$ 70 & 10.1-11.7 & 44.1 & Disk or Halo& $9.9 \times 10^{-4}$\\
 \hline
9020 & MS F0-F2& 2.7-3.6 & $5.2 \pm 1.0  \times 10^5$ & &  \\
9020 & Young WD  & 13.3-13.3 & $6.2  \pm  1.4 \times 10^3$& 0.5-1.6 & 793.9 &  \\
9020 & Old WD & 17.5-18.4 & 576 $\pm$ 123 & 10.7-12.9 & 73.7 & Disk or Halo & $3.16 \times 10^{-3}$\\
 \hline
9230 & MS K2-K3& 6.3-6.7& $6.3 \pm 0.4 \times 10^3$ & &  & K2-K3\\
9230 & Young WD  & 13.6-14.6 & 187 $\pm$ 41 & 1.1-3.2 & 23.9 & & \\
9230 & Old WD & 17.5-18.4 & 32 $\pm$ 7 & 10.4-12.3 & 4.0 & & \\
\enddata
\end{deluxetable}

\end{document}